\documentclass[twocolumn,twoside,slac_two]{revtex4}
\usepackage{graphicx}
\usepackage{fancyhdr}
\usepackage{amsmath}
\usepackage{epsfig}
\newcommand{\mbc}{M_{\rm bc}}
\newcommand{\de}{\Delta E}

\newcommand{\bdk}{$B^{\pm}\to DK^{\pm}$}
\newcommand{\bdkm}{$B^{-}\to DK^{-}$}
\newcommand{\bdkp}{$B^{+}\to DK^{+}$}

\newcommand{\bdsk}{$B^{\pm}\to D^{*}K^{\pm}$}

\newcommand{\bdskp}{$B^{+}\to D^{*}K^{+}$}

\newcommand{\bdks}{$B^{\pm}\to DK^{*\pm}$}

\newcommand{\bddsk}{$B^{\pm}\to D^{(*)}K^{\pm}$}

\newcommand{\bddsks}{$B^{\pm}\to D^{(*)}K^{(*)\pm}$}

\newcommand{\bdksnr}{$B^{\pm}\to DK^0_S\pi^{\pm}$}

\newcommand{\dsdpim}{$D^{*-}\to \overline{D}{}^0\pi^{-}$}

\newcommand{\dkpp}{$\overline{D}{}^0\to K^0_S\pi^+\pi^-$}
\newcommand{\dkkk}{$\overline{D}{}^0\to K^0_S K^+ K^-$}

\input babarsym 

\newcommand{\phm}{\ensuremath{\phantom{-}}} 
 
\newcommand{\phz}{\ensuremath{\phantom{0}}} 
\newcommand{\real}{\Re} 
\newcommand{\imag}{\Im}

\def\kskk{\ensuremath{\KS \Kp \Km}\xspace} 
\def\Dztokskk{\ensuremath{\Dz \to \kskk}\xspace}

\def \xbxbstmp {\ensuremath {x_\mp^{(\ast)}}\xspace}

\def \ybybstmp {\ensuremath {y_\mp^{(\ast)}}\xspace} 

\def \xsmp {\ensuremath {x_{s\mp}}\xspace}

\def \ysmp {\ensuremath {y_{s\mp}}\xspace}

\def\Dztilde   {\ensuremath {\tilde{D}^0}\xspace} 
\def\Dstarztilde   {\ensuremath {\tilde{D}^{\ast 0}}\xspace} 

\begin{document}

\title{Measurements of $\mathbf{\gamma/\phi_3}$}

\author{P. Krokovny}
\affiliation{IPNS, KEK, Tsukuba, Japan}

\begin{abstract}
This report summarizes the progress in measuring the 
angle $\gamma$ (or $\phi_3$) of the Unitarity Triangle. 

\end{abstract}

\maketitle

\thispagestyle{fancy}

\section{Introduction}

% UT measurements in general

Measurements of the Unitarity Triangle parameters allow one to search for 
New Physics effects at low energies. Most of such measurements are 
currently performed at $B$ factories --- the $e^+e^-$ machines operated 
with the center-of-mass energy around 10 GeV at the $\Upsilon(4S)$ resonance, 
which primarily decays to $B$ meson pairs. 

% Angles of the UT

One angle, $\phi_1$ (or $\beta$)\footnote{Two different notations of the 
Unitarity Triangle are used: $\alpha$, $\beta$, $\gamma$ or 
$\phi_2$, $\phi_1$ and $\phi_3$, respectively. The second option 
(adopted by Belle collaboration) will be used throughout the paper 
except for the case when BaBar results are discussed. }, 
has been measured with high precision
at the BaBar \cite{babar_phi1} and Belle~\cite{belle_phi1} experiments. 
The measurement of the angle $\phi_2/\alpha$ is more difficult due 
to theoretical uncertainties in the calculation of the penguin diagram 
contribution. Precise determination of the third angle, $\phi_3/\gamma$, 
is possible, {\it e.g. }, in the decays \bdk.
Although it requires a lot more data than for the other angles,
it is theoretically clean due to the absence of loop contributions.
In recent years, a lot of progress has been achieved in the
methods of the precise determination of the $\phi_2$ and $\phi_3$ angles.
This report summarizes the most recent progress in measuring the 
angle $\phi_3/\gamma$. 

\section{GLW analyses}

% Description of the method

\begin{figure}
\begin{center}
\includegraphics[width=0.45\textwidth]{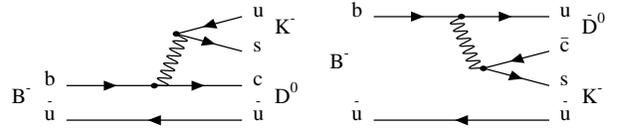}
\end{center}
\caption{Feynman diagrams for $B^{-} \rightarrow D^{0}K^{-}$ and 
$B^{-}\rightarrow\bar{D}^{0}K^{-}$.}
\label{feynman}
\end{figure}

The technique of measuring $\phi_3$ proposed by Gronau, London and Wyler
(and called GLW) \cite{glw} makes use of $D^0$ decays to CP eigenstates, 
such as $K^+K^-$, $\pi^+\pi^-$ (CP-even) or $K^0_S\pi^0$, $K^0_S\phi$
(CP-odd). Since both $D^0$ and $\bar{D}^{0}$
can decay into the same $CP$ eigenstate ($D_{CP}$, or $D_1$ for a 
$CP$-even state and $D_2$ for a $CP$-odd state),
the $b\rightarrow c$ and $b\rightarrow u$ processes shown in 
Fig.~\ref{feynman} interfere in the 
$B^{\pm} \to D_{CP} K^{\pm}$ decay channel. This interference may lead to
direct $CP$ violation. To measure $D$ meson decays to $CP$ eigenstates
a large number of $B$ meson decays are required since the branching fractions
to these modes are of order 1\%. To extract $\phi_3$ using the 
GLW method, the following observables sensitive to $CP$ violation are used: 
the asymmetries 
\begin{equation}
\label{glw_eq1}
\begin{split}
{\cal{A}}_{1,2} & \equiv \frac{{\cal B}(B^- \rightarrow D_{1,2}K^-) -
{\cal B}(B^+ \rightarrow D_{1,2}K^+) }{{\cal B}(B^- \rightarrow
D_{1,2}K^-) + {\cal B}(B^+ \rightarrow D_{1,2}K^+) }\\ 
& = \frac{2 r_B \sin \delta ' \sin \phi_3}{1 + r_B^2 + 2 r_B \cos \delta '
\cos \phi_3}
\end{split}
\end{equation}
and the double ratios 
\begin{equation}
\label{glw_eq2}
\begin{split}
{\cal{R}}_{1,2} &\equiv \frac{{\cal B}(B^- \rightarrow D_{1,2}K^-) +
{\cal B}(B^+ \rightarrow D_{1,2}K^+) }{{\cal B}(B^- \rightarrow
D^0 K^-) + {\cal B}(B^+ \rightarrow D^0 K^+) }\\
 &= 1 + r_B^2 + 2 r_B
\cos \delta ' \cos \phi_3, 
\end{split}
\end{equation}
where
\begin{equation}
\label{glw_eq3}
\delta ' = \left\{
             \begin{array}{ll}
              \delta_B & \mbox {{\rm  for }$D_1$}\\
              \delta_B + \pi&  \mbox{{\rm for }$D_2$}\\
             \end{array}
             \right. , 
\end{equation}
and $r_B \equiv |A(B^- \to \bar{D}^0 K^-)/A(B^- \to D^0 K^-)|$ 
is the ratio of the magnitudes of the two tree diagrams
shown in Fig.~\ref{feynman}, $\delta_B$ is their strong-phase
difference. 
The value of $r_B$ is given by the ratio of the CKM matrix elements 
$|V_{ub}^*V_{cs\vphantom{b}}^{\vphantom{*}}|/
 |V_{cb}^*V_{us\vphantom{b}}^{\vphantom{*}}|\sim 0.38$ 
and the color suppression factor.
Here we assume that mixing and $CP$ violation in the neutral $D$ meson
system can be neglected.

Instead of four observables $\mathcal{R}_{1,2}$ and $\mathcal{A}_{1,2}$, only three
of which are independent (since $\mathcal{A}_1\mathcal{R}_1 = 
-\mathcal{A}_2\mathcal{R}_2$), an alternative set of three parameters 
can be used: 
\begin{equation}
  \begin{split}
  x_{\pm} &= r_B\cos(\delta_B\pm \phi_3)\\
          &= \frac{\mathcal{R}_1(1+\mathcal{A}_1)-
                   \mathcal{R}_2(1+\mathcal{A}_2)}{4}, 
  \end{split}
\end{equation}
and 
\begin{equation}
  r_B^2 = \frac{\mathcal{R}_1+\mathcal{R}_2+2}{2}. 
\end{equation}
The use of these observables allows for a direct comparison with the 
methods involving Dalitz plot analyses of $D^0$ (see Section~\ref{dalitz}), 
where the same parameters $x_{\pm}$ are obtained. 

% New BaBar measurement

Measurements of $B\to D_{\rm CP}K$ decays have been performed by both the 
BaBar~\cite{babar} and Belle~\cite{belle} collaborations. 
Recently, BaBar updated their GLW 
analysis using the data sample of 382M $B\overline{B}$ pairs~\cite{babar_glw}. 
The analysis uses $D^0$ decays to $K^+K^-$ and $\pi^+\pi^-$ as $CP$-even 
modes, $K^0_S\pi^0$ and $K^0_S\omega$ as $CP$-odd modes. 

\begin{table}
  \caption{Results of the GLW analysis by BaBar}
  \label{glw_table}
  \begin{tabular}{|l|l|}
    \hline
    $\mathcal{R}_{1}$ & $\phantom{-}1.06\pm 0.10 \pm 0.05$ \\
    $\mathcal{R}_{2}$ & $\phantom{-}1.03\pm 0.10 \pm 0.05$ \\
    $\mathcal{A}_{1}$ & $+0.27\pm 0.09 \pm 0.04$ \\
    $\mathcal{A}_{2}$ & $-0.09\pm 0.09 \pm 0.02$ \\
    \hline
    $x_+$ & $-0.09\pm 0.05\pm 0.02$ \\
    $x_-$ & $+0.10\pm 0.05\pm 0.03$ \\
    $r_B^2$ & $0.05\pm 0.07\pm 0.03$ \\
    \hline
  \end{tabular}
\end{table}

The results of the analysis (both in terms of asymmetries and double 
ratios, and alternative $x_{\pm}, r^2_B$ set) are shown in Table~\ref{glw_table}. 
As follows from (\ref{glw_eq1}) and (\ref{glw_eq3}), the signs of the 
$\mathcal{A}_{1}$ and $\mathcal{A}_{2}$ asymmetries should be opposite, 
which is confirmed by the experiment. The $x_{\pm}$ values are in a
good agreement with the ones obtained by Dalitz analysis technique. 

\section{ADS analyses}

% Description of the method 

The difficulties in the application of the GLW methods arise primarily 
due to the small magnitude of the $CP$ asymmetry of the 
$B^{\pm}\to D_{CP}K^{\pm}$ decay probabilities, which may lead 
to significant systematic uncertainties in the observation of 
$CP$ violation. An alternative approach was proposed by Atwood, 
Dunietz and Soni \cite{ads}. Instead of using the $D^0$ decays to 
$CP$ eigenstates, the ADS method uses Cabibbo-favored and doubly 
Cabibbo-suppressed decays: $\overline{D}^0\to K^-\pi^+$ and
$D^0\to K^-\pi^+$. In the decays $B^+\to [K^-\pi^+]_D K^+$ and 
$B^-\to [K^+\pi^-]_D K^-$, the suppressed $B$ decay corresponds to the
Cabibbo-allowed $D^0$ decay, and vice versa. Therefore, the interfering 
amplitudes are of similar magnitudes, and one can expect the 
significant $CP$ asymmetry. 

Unfortunately, the branching ratios of the decays mentioned above 
are so small that they cannot be observed using the current experimental 
statistics. The observable that is measured in the ADS method is 
the fraction of  the suppressed and allowed branching ratios:
\begin{equation}
  \begin{split}
  \mathcal{R}_{ADS}&=\frac{Br(B^{\pm}\to [K^{\mp}\pi^{\pm}]_DK^{\pm})}
                   {Br(B^{\pm}\to [K^{\pm}\pi^{\mp}]_DK^{\pm})}\\
             &=r_B^2+r_D^2+2r_Br_D\cos\phi_3\cos\delta, 
  \end{split}
\end{equation}
where $r_D$ is the ratio of the doubly Cabibbo-suppressed and 
Cabibbo-allowed $D^0$ decay amplitudes:
\begin{equation}
  r_D=\left|\frac{A(D^0\to K^+\pi^-)}{A(D^0\to K^-\pi^+)}\right|=0.060\pm 0.002,
\end{equation}
and $\delta$ is a sum of strong phase differences in $B$ and $D$ decays: 
$\delta=\delta_B+\delta_D$. 

% New Belle measurement

The update of the ADS analysis using 657M $B\overline{B}$ pair was 
recently reported by Belle~\cite{belle_ads}. The analysis uses \bdk\ decays 
with $D^0$ decaying to $K^+\pi^-$ and $K^-\pi^+$ modes (and their 
charge-conjugated partners). The ratio of the suppressed and allowed 
modes is 
\begin{equation}
  \mathcal{R}_{ADS}=(8.0^{+6.3}_{-5.7}{}^{+2.0}_{-2.8})\times 10^{-3}. 
\end{equation}
Belle also reports the measurement of the $CP$ asymmetry, which appears 
to be consistent with zero:
\begin{equation}
  \mathcal{A}_{ADS}=-0.13^{+0.98}_{-0.88}\pm 0.26. 
\end{equation}
The ADS analysis currently does not give a significant constraint on $\phi_3$, 
but it provides important information on the value of $r_B$. 
Using the conservative assumption $\cos{\phi_3}\cos{\delta}=-1$ one 
obtains the upper limit $r_B<0.19$ at the 90\% CL. A somewhat tighter constraint 
can be obtained by using the $\phi_3$ and $\delta_B$ measurements from the 
Dalitz analyses (see Section~\ref{dalitz}), and the recent CLEO-c measurement 
of the strong phase 
$\delta_D=(22^{+11}_{-12}{}^{+9}_{-11})^{\circ}$~\cite{cleo_delta_d}. 

\section{Dalitz plot analyses}

\label{dalitz}

% Description of the method

A Dalitz plot analysis of a three-body final state of the $D$ meson
allows one to obtain all the information required for determination
of $\phi_3$ in a single decay mode. The use of a Dalitz plot analysis
for the extraction of $\phi_3$ was first discussed
by D. Atwood, I. Dunietz and A. Soni, in the context of the ADS
method \cite{ads}. This technique uses the interference of
Cabibbo-favored $D^0\to K^-\pi^+\pi^0$ and doubly Cabibbo-suppressed
$\overline{D}{}^0\to K^-\pi^+\pi^0$ decays.
However, the small rate for the doubly Cabibbo-suppressed decay
limits the sensitivity of this technique.

Three body final states such as 
$K^0_S\pi^+\pi^-$ \cite{giri,binp_dalitz} have been suggested as 
promising modes for the extraction of $\phi_3$. Like in the GLW or ADS 
method, the two amplitudes 
interfere as the $D^0$ and $\overline{D}{}^0$ mesons decay
into the same final state $K^0_S \pi^+ \pi^-$; 
we denote the admixed state as $\tilde{D}_+$. 
Assuming no $CP$ asymmetry in neutral $D$ decays, 
the amplitude of the $\tilde{D}_+$ decay 
as a function of Dalitz plot variables $m^2_+=m^2_{K^0_S\pi^+}$ and 
$m^2_-=m^2_{K^0_S\pi^-}$ is 
\begin{equation}
  f_{B^+}=f_D(m^2_+, m^2_-)+r_Be^{i\phi_3+i\delta_B}f_D(m^2_-, m^2_+), 
\end{equation}
where $f_D(m^2_+, m^2_-)$ is the amplitude of the \dkpp\ decay. 

Similarly, the amplitude of the $\tilde{D}_-$ decay from \bdkm\ process is
\begin{equation}
  f_{B^-}=f_D(m^2_-, m^2_+)+r_Be^{-i\phi_3+i\delta_B}f_D(m^2_+, m^2_-). 
\end{equation}
The \dkpp\ decay amplitude $f_D$ can be determined
from a large sample of flavor-tagged \dkpp\ decays 
produced in continuum $e^+e^-$ annihilation. Once $f_D$ is known, 
a simultaneous fit of $B^+$ and $B^-$ data allows the 
contributions of $r_B$, $\phi_3$ and $\delta_B$ to be separated. 
The method has only a two-fold ambiguity: 
$(\phi_3,\delta_B)$ and $(\phi_3+180^{\circ}, \delta_B+180^{\circ})$
solutions cannot be distinguished. 
References \cite{giri} and \cite{belle_phi3_prd} give  
a more detailed description of the technique. 

% Belle measurement: 

Both Belle and BaBar collaborations reported recently the updates of the 
$\phi_3(\gamma)$ measurements using a Dalitz plot analysis. The preliminary 
result obtained by Belle~\cite{belle_dalitz} uses the data sample of 657M 
$B\overline{B}$ pairs and two modes, \bdk\ and \bdsk\ with $D^{*}\to D\pi^0$. 
The neutral $D$ meson is reconstructed in $K^0_S\pi^+\pi^-$ final state 
in both cases. 

% Parametrization of the D0 decay amplitude

To determine the decay amplitude, $D^{*\pm}$ mesons
produced via the $e^+ e^-\to c\bar{c}$ continuum process are used, which 
then decay to a neutral $D$ and a charged pion. 
The flavor of the neutral $D$ meson is tagged by the charge of the pion 
in the decay \dsdpim. $B$ factories offer large sets of such charm data: 
$290.9\times 10^3$ events are used in Belle analysis with only 1.0\% background. 
 
The description of the \dkpp\ decay amplitude is based on the isobar model. 
The amplitude $f_D$ is represented by a coherent sum of two-body decay 
amplitudes and one non-resonant decay amplitude,
\begin{equation}
  f_D(m^2_+, m^2_-) = \sum\limits_{j=1}^{N} a_j e^{i\xi_j}
  \mathcal{A}_j(m^2_+, m^2_-)+
    a_{\text{NR}} e^{i\xi_{\text{NR}}}, 
  \label{d0_model}
\end{equation}
where $\mathcal{A}_j(m^2_+, m^2_-)$ is the matrix element, $a_j$ and 
$\xi_j$ are the amplitude and phase of the matrix element, respectively, 
of the $j$-th resonance, and $a_{\text{NR}}$ and $\xi_{\text{NR}}$
are the amplitude and phase of the non-resonant component. 
The model includes a set of 18 two-body amplitudes: 
five Cabibbo-allowed amplitudes: $K^*(892)^+\pi^-$, 
$K^*(1410)^+\pi^-$, $K_0^*(1430)^+\pi^-$, 
$K_2^*(1430)^+\pi^-$ and $K^*(1680)^+\pi^-$;  
their doubly Cabibbo-suppressed partners; eight amplitudes with
$K^0_S$ and a $\pi\pi$ resonance:
$K^0_S\rho$, $K^0_S\omega$, $K^0_Sf_0(980)$, $K^0_Sf_2(1270)$, 
$K^0_Sf_0(1370)$, $K^0_S\rho(1450)$, $K^0_S\sigma_1$ and $K^0_S\sigma_2$; and a 
flat non-resonant term. 
The free parameters of the fit are the amplitudes
$a_j$ and phases $\xi_j$ of the resonances, and 
the amplitude $a_{NR}$ and phase $\xi_{NR}$ of the non-resonant component. 
The results of the \dkpp\ amplitude fit are shown in Table~\ref{dkpp_table}. 

\begin{table}
\caption{Fit results for \dkpp\ decay. Errors are statistical only.}
\label{dkpp_table}
\begin{tabular}{|l|c|c|} \hline
Intermediate state           & Amplitude 
			     & Phase ($^{\circ}$) 
			     \\ \hline

$K_S \sigma_1$               & $1.56\pm 0.06$
                             & $214\pm 3$
                             \\

$K_S\rho^0$                  & $1.0$ (fixed)                                 
                             & 0 (fixed)   
                             \\

$K_S\omega$                  & $0.0343\pm 0.0008$
                             & $112.0\pm 1.3$
                             \\

$K_S f_0(980)$               & $0.385\pm 0.006$
                             & $207.3\pm 2.3$
                             \\

$K_S \sigma_2$               & $0.20\pm 0.02$
                             & $212\pm 12$
                             \\

$K_S f_2(1270)$              & $1.44\pm 0.04$
                             & $342.9\pm 1.7$
                             \\

$K_S f_0(1370)$              & $1.56\pm 0.12$ 
                             & $110\pm 4$
                             \\

$K_S \rho^0(1450)$           & $0.49\pm 0.08$
                             & $64\pm 11$
                             \\

$K^*(892)^+\pi^-$            & $1.638\pm 0.010$
                             & $133.2\pm 0.4$
                             \\ 

$K^*(892)^-\pi^+$            & $0.149\pm 0.004$
                             & $325.4\pm 1.3$
                             \\

$K^*(1410)^+\pi^-$	     & $0.65\pm 0.05$
			     & $120\pm 4$
			     \\

$K^*(1410)^-\pi^+$	     & $0.42\pm 0.04$
			     & $253\pm 5$
			     \\

$K_0^*(1430)^+\pi^-$         & $2.21\pm 0.04$
                             & $358.9\pm 1.1$
                             \\

$K_0^*(1430)^-\pi^+$         & $0.36\pm 0.03$
                             & $87\pm 4$
                             \\

$K_2^*(1430)^+\pi^-$         & $0.89\pm 0.03$
                             & $314.8\pm 1.1$
                             \\

$K_2^*(1430)^-\pi^+$         & $0.23\pm 0.02$
                             & $275\pm 6$
                             \\

$K^*(1680)^+\pi^-$           & $0.88\pm 0.27$
                             & $82\pm 17$
                             \\

$K^*(1680)^-\pi^+$           & $2.1\pm 0.2$
                             & $130\pm 6$
                             \\

non-resonant                 & $2.7\pm 0.3$
                             & $160\pm 5$
                             \\ 
\hline
\end{tabular}
\end{table}

% Signal yields

The selection of \bddsk\ decays is based on the CM energy difference
$\Delta E = \sum E_i - E_{\rm beam}$ and the beam-constrained $B$ meson mass
$M_{\rm bc} = \sqrt{E_{\rm beam}^2 - (\sum \vec{p}_i)^2}$, where $E_{\rm beam}$ 
is the CM beam 
energy, and $E_i$ and $\vec{p}_i$ are the CM energies and momenta of the
$B$ candidate decay products. To suppress background from 
$e^+e^-\to q\bar{q}$ ($q=u, d, s, c$) continuum events, the variables 
that characterize the event shape are also calculated. At the first stage of 
the analysis, when the $(\mbc, \de)$ distribution is fitted in order to 
obtain the fractions of the background components, the requirement on the 
event shape is imposed to suppress the continuum events. The number of 
such ``clean" events is 756 for \bdk\ mode with 29\% background, and 
149 events for \bdsk\ mode with 20\% background. 
In the Dalitz plot fit, the events are not rejected based on event 
shape variables, these are used in the likelihood function to 
better separate signal and background events. 

% Fit results

The Dalitz distributions of the 
$B^+$ and $B^-$ samples are fitted separately, using Cartesian parameters 
$x_{\pm}=r_{\pm}\cos(\pm\phi_3+\delta_B)$ and 
$y_{\pm}=r_{\pm}\sin(\pm\phi_3+\delta_B)$, where the indices ``$+$" and 
``$-$" correspond to $B^+$ and $B^-$ decays, respectively. 
In this approach the amplitude ratios ($r_+$ and $r_-$) are 
not constrained to be equal for the $B^+$ and $B^-$ samples. 
Confidence intervals in $r_B$, $\phi_3$ and $\delta_B$ are then obtained 
from the $(x_{\pm},y_{\pm})$ using a frequentist technique. 
The values of the fit parameters $x_{\pm}$ and $y_{\pm}$ are 
listed in Table~\ref{sig_fit_table}. 

\begin{table}
  \caption{Results of the signal fits in parameters $(x,y)$. The first error 
  is statistical, the second is experimental systematic error. 
  Model uncertainty is not included. }
  \label{sig_fit_table}
  \begin{tabular}{|l|c|c|}
  \hline
  Parameter  & \bdkp & \bdskp \\ 
  \hline
  $x_-$ & $+0.105\pm 0.047\pm 0.011$ & $+0.024\pm 0.140\pm 0.018$ \\
  $y_-$ & $+0.177\pm 0.060\pm 0.018$ & $-0.243\pm 0.137\pm 0.022$ \\
  $x_+$ & $-0.107\pm 0.043\pm 0.011$ & $+0.133\pm 0.083\pm 0.018$ \\
  $y_+$ & $-0.067\pm 0.059\pm 0.018$ & $+0.130\pm 0.120\pm 0.022$ \\ 
  \hline
  \end{tabular}
\end{table}

The values of the 
parameters $r_B$, $\phi_3$ and $\delta_B$ obtained from the combination of 
\bdk\ and \bdsk\ modes are presented in Table~\ref{fc_comb_table}. 
Note that in addition to the detector-related systematic error which 
is caused by the uncertainties of the background description, 
imperfect simulation etc., the result suffers from the uncertainty of 
the $D$ decay amplitude description. 
The statistical confidence level of $CP$ violation for the combined result
is $(1-5.5\times 10^{-4})$, or 3.5 standard deviations.

%\begin{table*}
%  \caption{$CP$ fit results. The first error is statistical, the second 
%  is experimental systematic, and the third is the $D^0$ model uncertainty. }
%  \label{fit_res_table}
%  \begin{tabular}{|l||c|c|} \hline
%  Parameter & \bdkp\ mode
%            & \bdskp\ mode\\ \hline
%  $\phi_3$  & $80.8^{\circ}\;^{+13.1^{\circ}}_{-14.8^{\circ}}\pm 5.0^{\circ}\pm 8.7^{\circ}$ 
%            & $63.8^{\circ}\;^{+20.8^{\circ}}_{-22.9^{\circ}}\pm 4.7^{\circ}\pm 8.7^{\circ}$ 
%            \\
%  $r$       & $0.161^{+0.040}_{-0.038}\pm 0.011\pm 0.049$
%            & $0.208^{+0.085}_{-0.083}\pm 0.015\pm 0.049$
%            \\
%  $\delta$  & $137.4^{\circ}\;^{+13.0^{\circ}}_{-15.7^{\circ}}\pm 4.0^{\circ}\pm 22.9^{\circ}$
%            & $342.0^{\circ}\;^{+21.4^{\circ}}_{-22.9^{\circ}}\pm 3.7^{\circ}\pm 22.9^{\circ}$
%            \\
%  \hline
%  \end{tabular}
%\end{table*}

\begin{table*}
  \caption{Results of the combination of \bdkp\ and \bdskp\ modes. }
  \label{fc_comb_table}
  \begin{tabular}{|l|c|c|c|c|c|} \hline
  Parameter & $1\sigma$ interval & $2\sigma$ interval & 
              Systematic error & Model uncertainty \\ \hline
  $\phi_3$  & $76^{\circ}\;^{+12^{\circ}}_{-13^{\circ}}$ 
            & $49^{\circ}<\phi_3<99^{\circ}$ 
            & $4^{\circ}$ & $9^{\circ}$ \\
  $r_{DK}$  & $0.16\pm 0.04$ 
            & $0.08<r_{DK}<0.24$ 
            & $0.01$ & $0.05$ \\
  $r_{D^*K}$  & $0.21\pm 0.08$ 
            & $0.05<r_{D^*K}<0.39$ 
            & $0.02$ & $0.05$ \\
  $\delta_{DK}$  & $136^{\circ}\;^{+14^{\circ}}_{-16^{\circ}}$ 
            & $100^{\circ}<\delta_{DK}<163^{\circ}$ 
            & $4^{\circ}$ & $23^{\circ}$ \\
  $\delta_{D^*K}$  & $343^{\circ}\;^{+20^{\circ}}_{-22^{\circ}}$ 
            & $293^{\circ}<\delta_{DK}<389^{\circ}$ 
            & $4^{\circ}$ & $23^{\circ}$ \\
  \hline
  \end{tabular}
\end{table*}
% BaBar measurement: 

In contrast to the Belle analysis, BaBar~\cite{babar_dalitz} uses a smaller 
data sample of 383M $B\overline{B}$ pairs, but analyses seven different 
decay modes: \bdk, \bdsk\ with $D^0\to D\pi^0$ and $D\gamma$, and \bdks, 
where the neutral $D$ meson is reconstructed in $K^0_S\pi^+\pi^-$ 
and $K^0_SK^+K^-$ (except for \bdks\ mode) final states. The signal 
yields for these modes are shown in Table~\ref{babar_yield}. 

\begin{table}
  \caption{Signal yields of different modes used for Dalitz analysis by 
            BaBar collaboration~\cite{babar_dalitz}. }
  \label{babar_yield}
  \begin{tabular}{|l|l|l|}
    \hline
    $B$ decay & $D$ decay & Yield \\
    \hline
    \bdk\                             & \dkpp    & $600\pm 31$ \\
                                       & \dkkk    & $112\pm 13$ \\
    $B^{\pm}\to [D\pi^0]_{D^{*}}K^{\pm}$ & \dkpp   & $133\pm 15$ \\
                                       & \dkkk    & $32\pm 7$ \\
    $B^{\pm}\to [D\gamma]_{D^{*}}K^{\pm}$ & \dkpp   & $129\pm 16$ \\
                                        & \dkkk    & $21\pm 7$ \\
    $B^{\pm}\to DK^{*\pm}$              & \dkpp   & $118\pm 18$ \\
                                      
    \hline
  \end{tabular}
\end{table}

% D0->KsPiPi and D0->KsKK amplitudes

The differences from the Belle model of \dkpp\ decay are as follows: 
the K-matrix formalism is used by default to describe the $\pi\pi$ $S$-wave, 
while the $K\pi$ $S$-wave is parametrized using $K^*_0(1430)$ resonances 
and an effective range non-resonant component with a phase shift. 

The description of \dkkk\ decay amplitude uses an isobar model with eight 
two-body decays: $K^0_S a_0(980)^0$, $K^0_S \phi(1020)$, 
$K^0_S f_0(1370)$, $K^0_S f_2(1270)^0$, $K^0_S a_0(1450)^0$, 
$K^- a_0(980)^+$, $K^+ a_0(980)^-$, and $K^- a_0(1450)^+$. The results 
of the \dkkk\ amplitude fit are shown in Table~\ref{kskk_model}. 

\begin{table}[hbt!]
\caption{\label{kskk_model} \CP eigenstates, CA, and DCS complex amplitudes $a_r e^{i\phi_r}$ and fit fractions,  
obtained from the fit of the \Dztokskk Dalitz plot distribution from $\Dstarp \to \Dz \pip$.  
Errors are statistical only.  
} 
%\begin{ruledtabular}
\begin{tabular}{|l|c|c|c|}
\hline
%\\[-0.15in]
    Component  &  $\real\{a_r e^{i\phi_r}\}$  &  $\imag\{a_r e^{i\phi_r}\}$ & Frac. (\%) \\ [0.01in] 
\hline 
$\KS a_0(980)^0$      &  $\phm1$                    &  $\phm0$                 & $55.8$  \\ 
$\KS \phi(1020)$      &  $-0.126\pm0.003$           &  $\phm0.189\pm0.005$     &  $44.9$   \\ 
$\KS f_0(1370)$       &  $-0.04\phz\pm0.06\phz$     &  $-0.00\phz\pm0.05\phz$  &  $\phz0.1$  \\ 
$\KS f_2(1270)$       &  $\phm0.257\pm0.019$        &  $-0.041\pm0.026$        &  $\phz0.3$      \\ 
$\KS a_0(1450)^0$     &  $\phm0.06\phz\pm0.12\phz$  &  $-0.65\phz\pm0.09\phz$  & $12.6$       \\ 
\hline 
$K^- a_0(980)^+$      & $-0.561\pm0.015$            &  $\phm0.01\phz\pm0.03\phz$ &  $16.0$  \\ 
$K^- a_0(1450)^+$     & $-0.11\phz\pm0.06\phz$      &  $\phm0.81\phz\pm0.03\phz$ &  $21.8$  \\ 
\hline 
$K^+ a_0(980)^-$      &  $-0.087\pm0.016$           & $\phm0.079\pm0.014$        &  $\phz0.7$       \\ 
\hline
\end{tabular}
%\end{ruledtabular}
\end{table}

% Fit results

The fit to signal samples is performed in a similar way as in Belle analysis, 
using the unbinned likelihood function that includes Dalitz plot variables, 
$B$ meson selection variables, and event shape parameters. The results of the 
fit in Cartesian parameters are shown in Table~\ref{babar_cp_coord}. 
In the combination of all modes, BaBar obtains 
$\gamma=(76^{+23}_{-24}\pm 5\pm 5)^{\circ}$ (mod 180$^{\circ}$). 
The values of the amplitude ratios are $r_B=0.086\pm 0.035\pm 0.010\pm 0.011$
for \bdk, $r^*_B=0.135\pm 0.051\pm 0.011\pm 0.005$ for \bdsk, and 
$\kappa r_s=0.163^{+0.088}_{-0.105}\pm 0.037\pm 0.021$
for \bdks\ (here $\kappa$ accounts for possible nonresonant \bdksnr\ 
contribution). The significance of the direct $CP$ violation is 
99.7\%, or 3.0 standard deviations. 

\begin{table*}[hbt!] 
\caption{\label{babar_cp_coord} \CP-violating parameters \xbxbstmp, \ybybstmp,
  \xsmp, and \ysmp.
%, as obtained from the fit.  
The first error is statistical, the second is experimental systematic uncertainty and the third is  
the systematic uncertainty associated with the Dalitz models.} 
\begin{tabular}{|l|c|c|c|}
\hline
Parameters                   & $\Bm \to \Dztilde \Km$ & $\Bm \to \Dstarztilde \Km$ & $\Bm \to \Dztilde \Kstarm$ \\ [0.01in] \hline 
$x_{-}~,~x_{-}^{*}~,~x_{s-}$ & $\phm0.090\pm 0.043\pm 0.015 \pm 0.011$ & $-0.111\pm 0.069\pm 0.014\pm 0.004$    & $\phm0.115\pm0.138\pm0.039\pm0.014$\\ 
$y_{-}~,~y_{-}^{*}~,~y_{s-}$ & $\phm0.053\pm 0.056\pm 0.007 \pm 0.015$ & $-0.051\pm 0.080\pm 0.009\pm 0.010$    & $\phm0.226\pm0.142\pm0.058\pm0.011$\\ 
$x_{+}~,~x_{+}^{*}~,~x_{s+}$ & $-0.067\pm 0.043 \pm 0.014\pm 0.011$    & $\phm0.137\pm 0.068\pm 0.014\pm 0.005$ & $-0.113\pm0.107\pm0.028\pm0.018$ \\ 
$y_{+}~,~y_{+}^{*}~,~y_{s+}$ & $-0.015\pm 0.055\pm 0.006\pm 0.008$     & $\phm0.080\pm 0.102\pm 0.010\pm 0.012$ & $\phm0.125\pm0.139\pm0.051\pm0.010$\\ 
\hline
\end{tabular} 
\end{table*} 

\section{Other techniques}

% BaBar: B0 -> D0 K*0 decay

Several other decays involving neutral $B$ mesons have been tried by 
the BaBar collaboration for $\gamma$ measurement. One of them is the decay
$B^0\to DK^*(892)^0$, where the similar Dalitz analysis of the 
three-body decay \dkpp\ is performed. Similarly to \bdk\ decays, 
this mode allows for direct measurement of the angle $\gamma$, but 
since both amplitudes involving $D^0$ and $\overline{D}{}^0$ are 
color-suppressed, the value of $r_B$
is larger $r_B\sim 0.4$. The flavor of the $B$ meson is tagged by the 
charges of the $K^*(892)^0$ decay products ($K^+\pi^-$ or $K^-\pi^+$). 

The analysis based on 371M $B\overline{B}$ pairs was performed~\cite{bndks}. 
The analysis procedure is similar to that with 
charged $B$ mesons. The fit yields the following constraints on 
$\gamma$ and amplitude ratio $r_B$: 
$\gamma=(162\pm 56)^{\circ}$, $r_B<0.55$ with 90\% CL. 

% BaBar: B0 -> D+- K0 pi-+ decay

Another neutral $B$ decay mode investigated by BaBar is 
$B^0\to D^{\mp}K^0\pi^{\pm}$. Similarly to measurements based on 
$B^0\to D^{(*)}\pi$ decays~\cite{bndspi_babar,bndspi_belle}, the interference between the 
$b\to u$ and $b\to c$ diagrams is achieved due to the mixing of 
neutral $B$ mesons. Therefore, this method requires to tag the flavor 
of the other $B$ meson and to perform a time-dependent analysis. 
As a result, this method is sensitive to the combination $2\beta+\gamma$
of the CKM angles~\cite{bdkpi,bdkpi2}. 

First advantage of this technique compared to the methods based 
on $B^0\to D^{(*)}\pi$ decays is that, since both 
$b\to c$ and $b\to u$ diagrams involved in this decay are color-suppressed, 
the expected value of the ratio $r$ is of the order 0.3. 
Secondly, $2\beta+\gamma$ is measured with only a two-fold ambiguity
(compared to four-fold in $B^0\to D^{(*)}\pi$ decays). In addition, 
all strong amplitudes and phases can be, in principle, measured in the 
same data sample. 

The BaBar collaboration has performed the analysis based on 347M 
$B\overline{B}$ pairs data sample~\cite{babar_bdkpi}. Time-dependent Dalitz 
plot analysis of the decay $B^0\to D^{\mp}K^0\pi^{\pm}$ is performed. 
This decay is found to be dominated by $B^0\to D^{**0}K^0_S$ 
(both $b\to u$ and $b\to c$ transitions) and $B^0\to D^-K^{*+}$ ($b\to c$) 
states. The analysis finds $558\pm 34$ flavor-tagged signal events, 
from the unbinned maximum likelihood fit to the time-dependent 
Dalitz distribution, the central value of the $2\beta+\gamma$ as a 
function of $r$ is obtained. The value of $r$ cannot be fixed with the 
current data sample, therefore, the value $r=0.3$ is used, and its error
is taken into account in the systematic error. This results in the 
value $2\beta+\gamma=(83\pm 53\pm 20)^{\circ}$ or $(263\pm 53\pm 20)^{\circ}$. 

\section{World average results}

% UTfit averages

\begin{figure}
  \epsfig{file=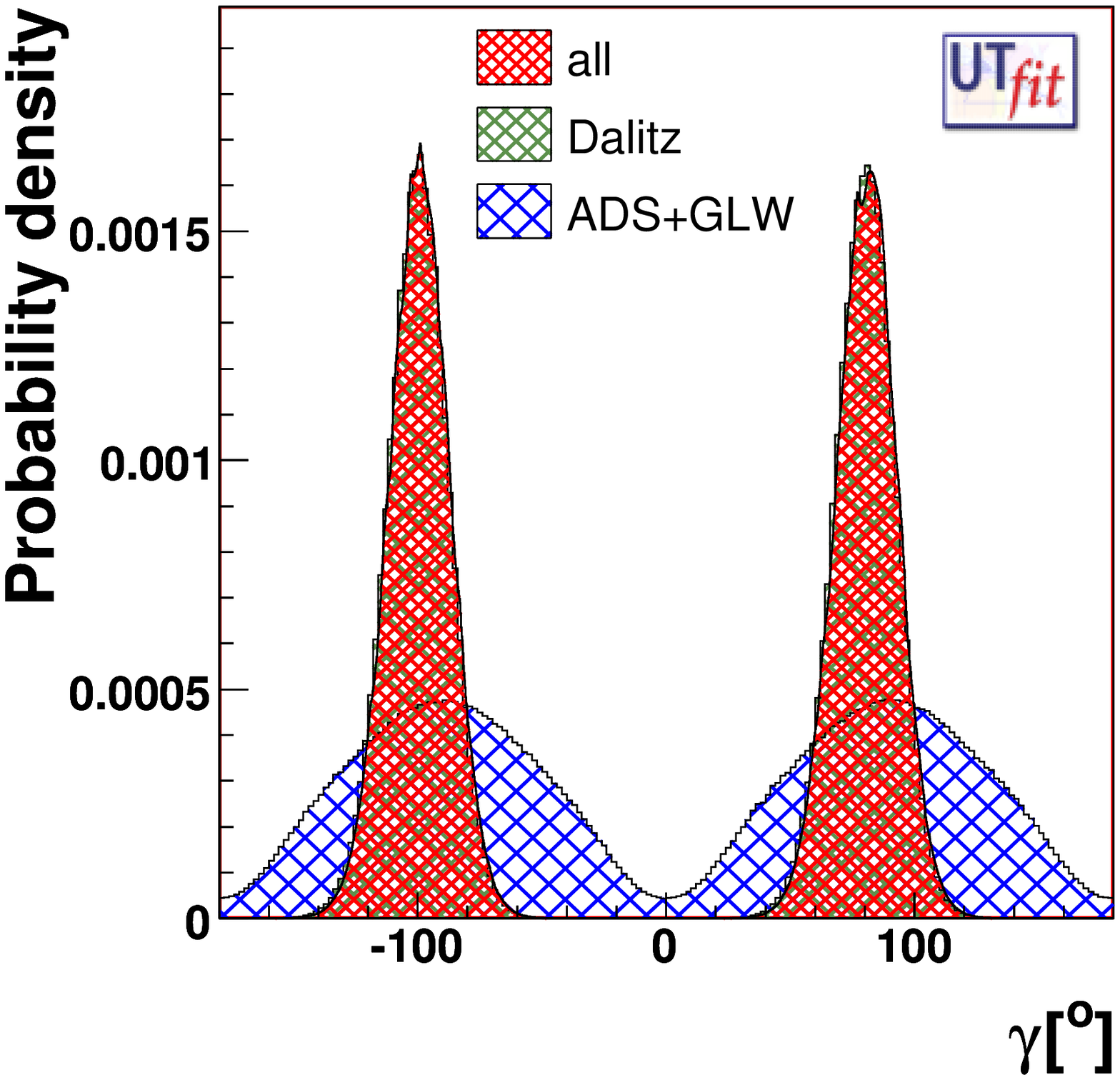,width=0.235\textwidth}
  \hfill
  \epsfig{file=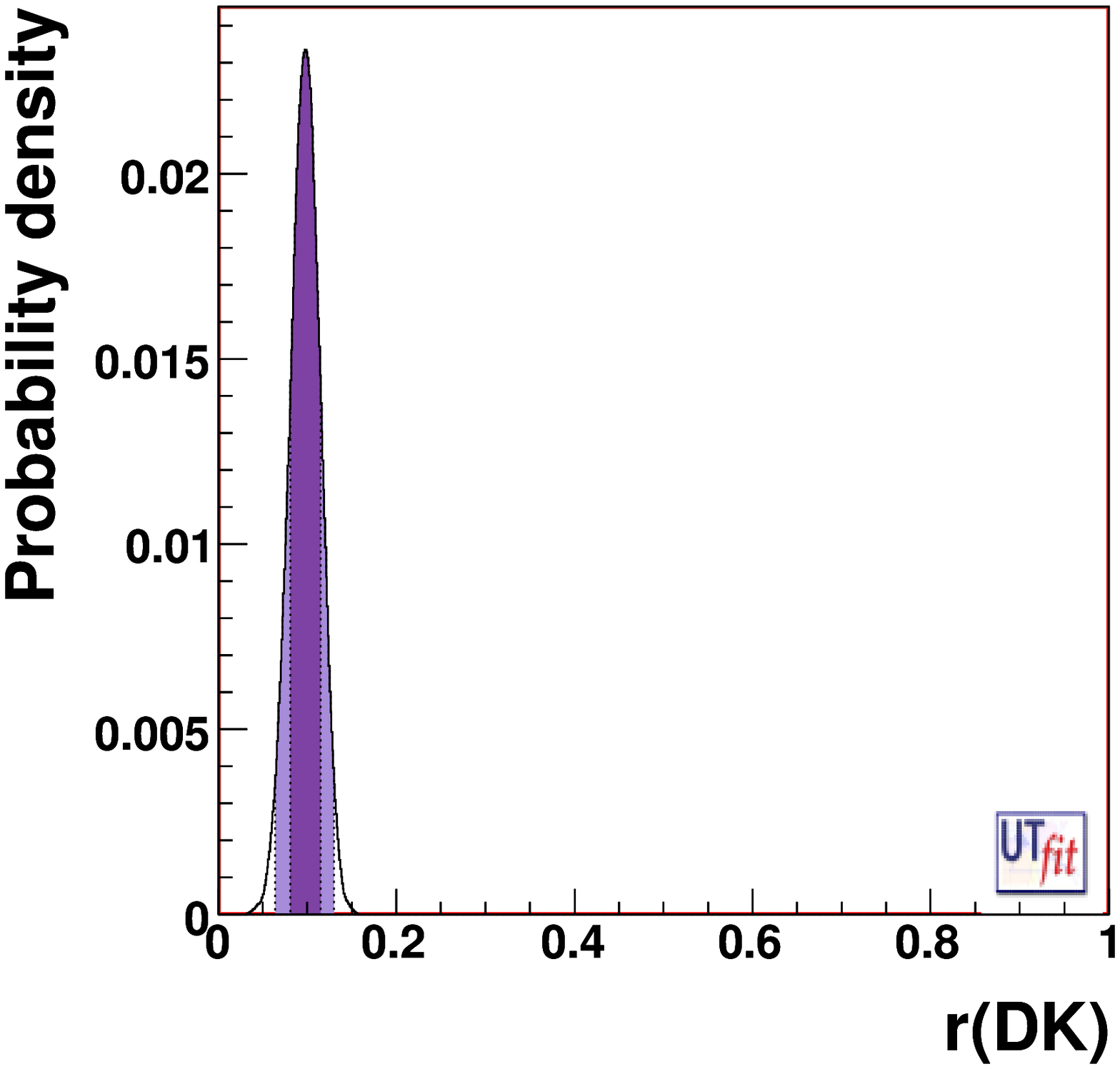,width=0.235\textwidth}

  \epsfig{file=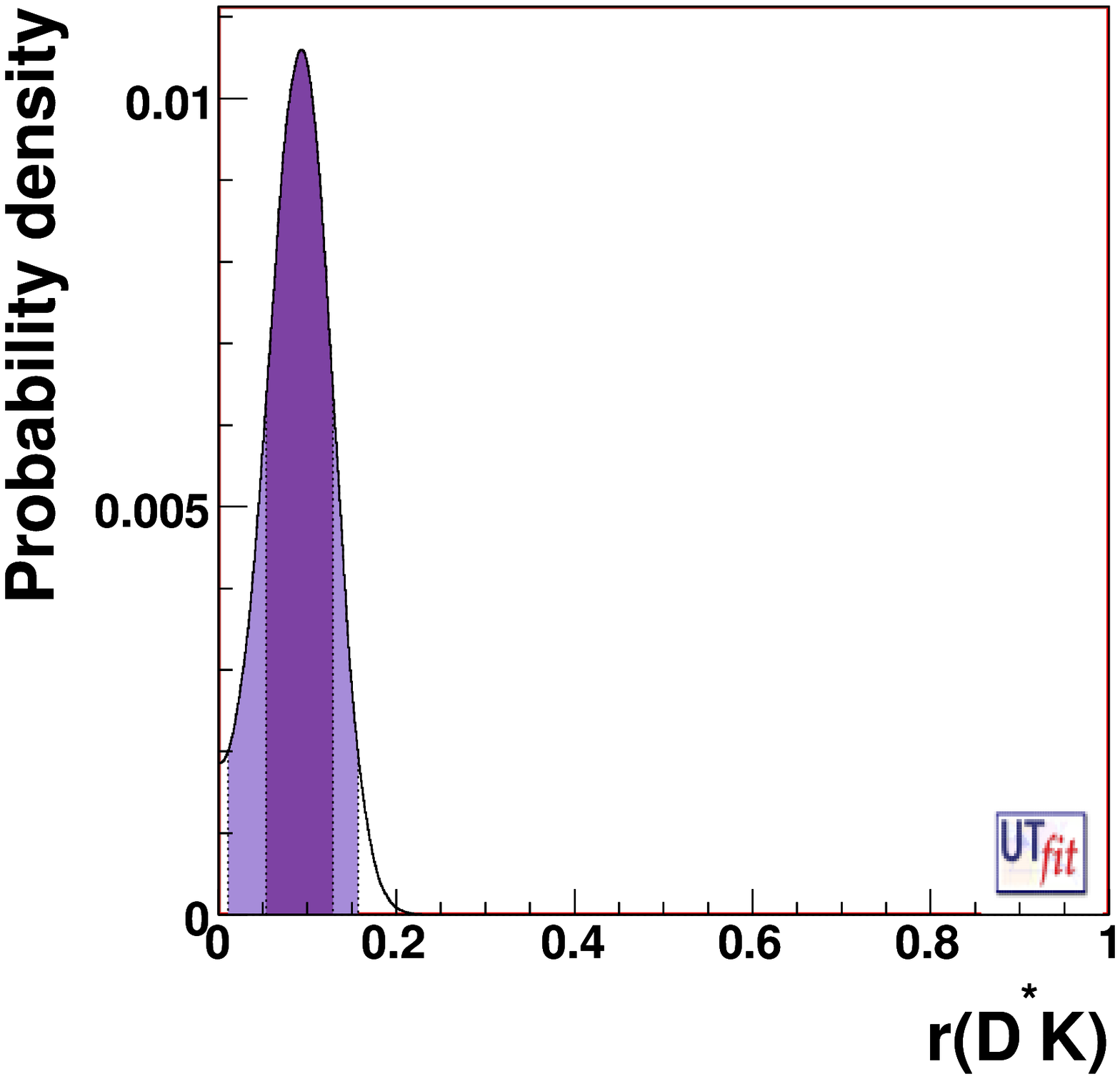,width=0.235\textwidth}
  \hfill
  \epsfig{file=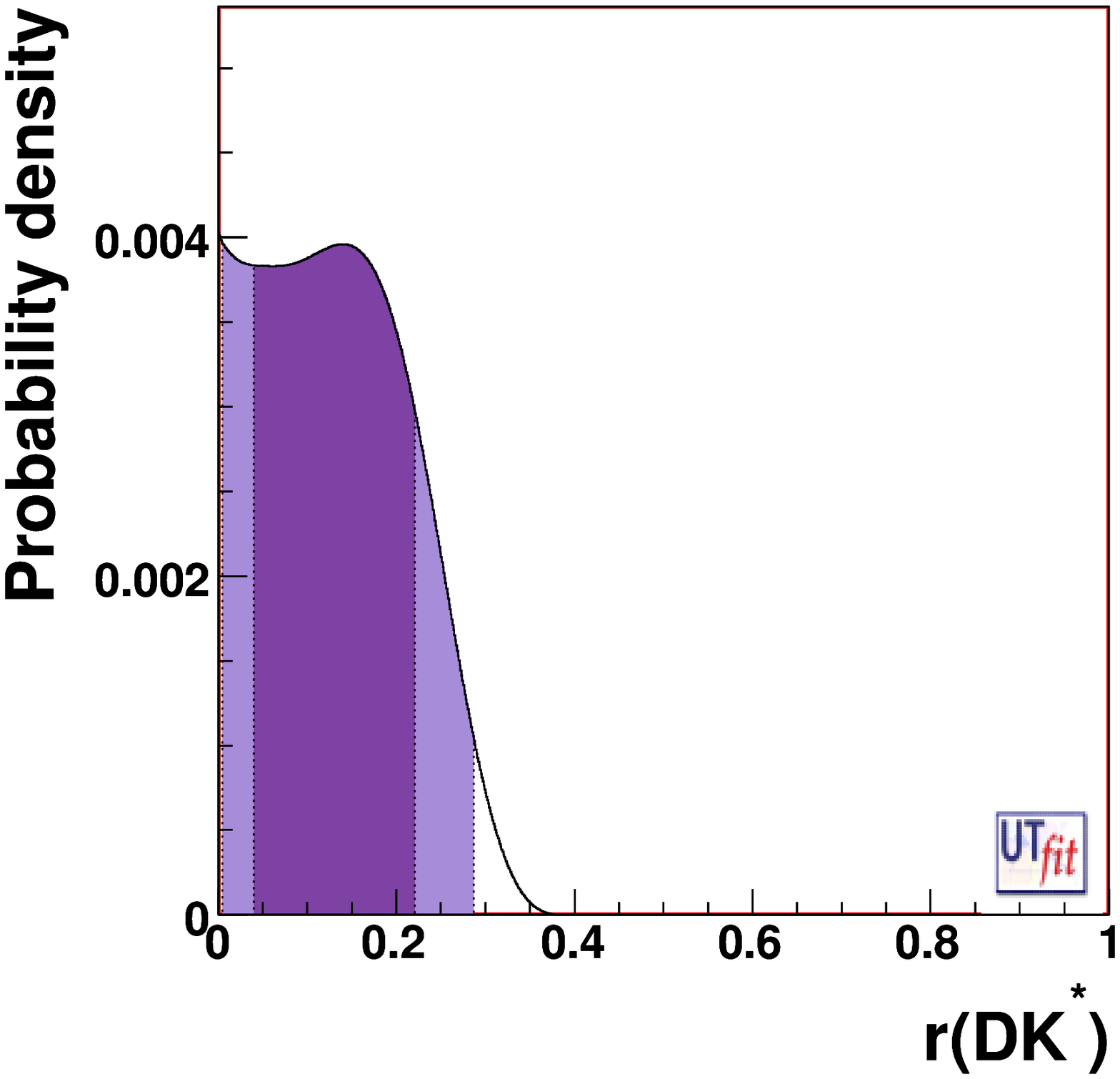,width=0.235\textwidth}
  \caption{Probability density functions for $\phi_3/\gamma$ (top left), 
            and $r_B$ in \bdk\ (top right), \bdsk\ (bottom left), and 
            \bdks\ (bottom right) based on all available $\phi_3/\gamma$ 
            measurements. }
  \label{utfit_pdf}
\end{figure}

The world average $\phi_3$ results that include the latest measurements 
presented in 2008, are available from the UTfit group~\cite{utfit}. 
The probability density functions for $\gamma$ and amplitude ratios $r_B$
are shown in Fig.~\ref{utfit_pdf}. The world average values for these 
parameters are $\phi_3/\gamma=(81\pm 13)^{\circ}$, $r_B(DK)=0.098\pm 0.017$, 
$r_B(D^*K)=0.092\pm 0.038$, $r_B(DK^*)=0.13\pm 0.09$. 

% Discussion of rB value and precision

Essential is the fact that for the first time the value of $r_B$
is shown to be significantly non-zero. In previous measurements, poor
$r_B$ constraint caused sufficiently non-gaussian errors for $\phi_3$, 
and made it difficult to predict the future sensitivity of this 
parameter. Now that $r_B$ is constrained to be of the order 0.1, 
one can confidently extrapolate the current precision to future measurements 
at LHCb and Super-B facilities. 

As it can be seen from Fig.~\ref{utfit_pdf} (top left), the $\phi_3/\gamma$
precision is mainly dominated by Dalitz analyses. These analyses have 
currently a hard-to-control uncertainty due to $D^0$ decay amplitude 
description, which is estimated to be 5--10$^{\circ}$. At the current level 
of statistical precision this error starts to influence the total 
$\phi_3/\gamma$ uncertainty. A solution to this problem can be the 
use of quantum-correlated $D\overline{D}$ decays at $\psi(3770)$ 
resonance available currently at CLEO-c experiment, where the 
missing information about the strong phase in $D^0$ decay can 
be obtained experimentally~\cite{modind,modind2}. With CLEO-c data sample,  
the $\gamma$ uncertainty due to $D$ decay amplitude can be as low 
as $5^{\circ}$ (and, since it becomes a statistical uncertainty, it 
is more reliable than the current estimation based on the arbitrary 
variations of the model), while with the future BES-III sample it 
can be lowered to a degree level. 

\begin{figure}
  \epsfig{file=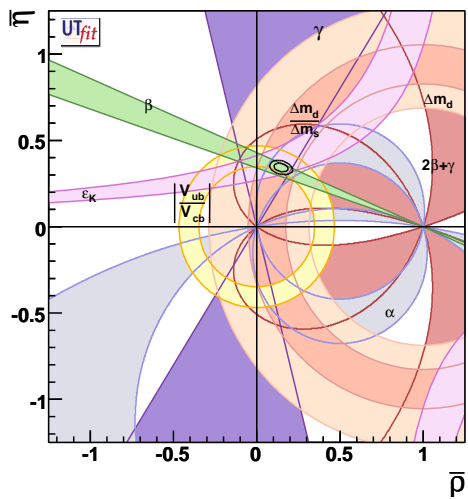,width=0.44\textwidth}
  \caption{UTfit constraints on the Unitarity Triangle vertex including 
  the latest $\phi_3/\gamma$ measurements. }
  \label{utfit_tri}
\end{figure}

UTfit constraints on the Unitarity Triangle vertex are shown 
in Fig.~\ref{utfit_tri}. The plot shows a good agreement between the 
different measurements, and $\phi_3/\gamma$ results, although still 
have poorer sensitivity compared to other angles measurements, 
fit well into the whole picture. 

\section{Conclusion}

In the past year, many new measurements related to a determination of 
$\phi_3/\gamma$ have appeared. As a result, strong evidence of a 
direct $CP$ violation in \bdk\ decays is obtained for the first time 
in a combination of B-factories results. Essential is that the amplitude 
ratio $r_B$, which determines the magnitude of the $CP$ violation and 
the precision of the $\phi_3/\gamma$ measurement, is finally constrained 
to be non-zero ($r_B=0.10\pm 0.02$ in the UTfit world average). This 
allows one to confidently extrapolate the sensitivity of $\phi_3/\gamma$
measurements to future experiments. Current world average is 
$\phi_3/\gamma=(81\pm 13)^{\circ}$; this value is dominated by the 
measurements based on Dalitz plot analyses of $D$ decay from \bddsks\ 
precesses. Although these analyses currently include a hard-to-control 
uncertainty due to the $D$ decay model, there are ways of dealing 
with this problem using charm data samples from CLEO-c and 
BES-III facilities, that should allow for a degree-level precision 
of $\phi_3/\gamma$ to be reached at the next generation $B$ factories. 

\bigskip % extra skip inserted

\end{document}